\documentclass[amsmath,amssymb,pra,twocolumn,showpacs]{revtex4}
\usepackage[cp1251]{inputenc}
\usepackage[english]{babel}
\usepackage{color}
\usepackage{graphicx}
\usepackage{dcolumn}
\usepackage{braket}
\usepackage{amsmath}
\usepackage{natbib}
\numberwithin{equation}{section}

\begin{document}
\bibliographystyle{unsrt}

\title{Invariant mass and propagation speed of light pulses in vacuum}


\author{M.V. Fedorov$^{1, 2, 3}$}
\email{fedorovmv@gmail.com}
\author{S.V. Vintskevich$^{1, 2}$}
\email{vintskevich@phystech.edu}
\address{$^1$A.M. Prokhorov General Physics Institute, Russian Academy of Sciences,Moscow,Russia \\
$^2$Moscow Institute of Physics and Technology, Dolgoprudny, Moscow Region, Russia\\
$^3$National Research Nuclear University MEPhI
(Moscow Engineering Physics Institute), 31 Kashirskoe Shosse, Moscow, 115409, Russia}

\date{\today}

\begin{abstract}
We show that the concept of the Lorentz-invariant mass of groups of particles can be applied to light pulses consisting of very large but finite numbers of noncollinear photons. Explicit expressions are found for the invariant mass of this manifold of photons for the case of diverging Gaussian light pulses propagating in vacuum. As the found mass is finite, the light pulses propagate in vacuum with a speed somewhat smaller than the light speed. A small difference between the light speed and the beam-propagation velocity is found to be related directly with the invariant mass of a pulse. Focusing and/or defocusing of light pulses is shown to strengthen the effect of pulse slowing down accompanied by increasing of the pulse invariant mass. A scheme for measuring these quantities experimentally is proposed and discussed

\end{abstract}

\maketitle
\bibliographystyle{unsrt}

\section{Introduction}
In the relativistic physics, the Lorentz-invariant mass of any object is known \cite{Landau} to be defined as its squared 4-momentum $p^{(4)}=\{\varepsilon/c,\,{\vec p}\}$ with ${\vec p}$  and $\varepsilon$ being the object's 3D momentum and energy,
\begin{equation}
 \label{mass1}
  m^2c^4=c^2{p^{(4)}}^2=\varepsilon^2-c^2{\vec{p}}^{\;2},
\end{equation}
and the velocity of an object $v$ given by
\begin{equation}
 \label{velocity}
 v=\frac{c^2\braket{p_z}}{\braket{\varepsilon}}.
\end{equation}
These definitions are valid for both massive objects and massless particles. If $m=0$, Eqs. (\ref{mass1}) and (\ref{velocity}) give $\varepsilon=c|{\vec p}|$ and $v=c$, i.e. the massless particles move with the speed of light. Oppositely, if $m\neq 0$, then $c|{\vec p}|<\varepsilon$ and $v<c$, i.e. the velocity of motion of massive objects and particles is always smaller than the speed of light.

The known massless particles are photons and for a single photon Eq. (\ref{mass1}) gives immediately $m=0$ as the energy and momentum of a photon are given by $\varepsilon=\hbar\omega$ and $|{\vec p}\,|=\hbar\omega/c$, where $\omega$ is the photon frequency.

The definition (\ref{mass1}) is easily generalized for groups of particles or objects because energies and momenta of particles are additive to give \cite{Okun,Okun1,Okun2,Rivlin}
\begin{equation}
  \label{mass-general}
  m^2c^4=\Big(\sum_{i}\varepsilon_i\Big)^2-c^2\Big(\sum_{i}{\vec p}_i\Big)^2\equiv c^2\sum_{i,j}\Big(p_i^{(4)}\cdot p_j^{(4)}\Big),
\end{equation}
where $\Big(p_i^{(4)}\cdot p_j^{(4)}\Big)$ are the Lorentz-invariant scalar products of the four-momenta of particles.

The simplest example is the pair of photons having, e.g., equal frequencies, $\omega_1=\omega_2=\omega$. If their wave vectors ${\vec k}_1$ and ${\vec k}_2$ are parallel to each other, their sum equals the double single-photon wave vector $2{\vec k}_1$ with the absolute value $2\omega/c$, and Eq. ({\ref{mass-general}}) gives the same result as for a single photon: $m_{1+1}=0$. If however, the wave vectors are not parallel, and the angle between them is $2\vartheta\neq 0$, the vectorial sum of wave vectors gives $|{\vec k}_1+{\vec k}_2|=2\cos\vartheta\,\omega/c\neq 2\omega/c$ and, hence,  Eq. (\ref{mass-general}) yields
\begin{equation}
  \label{2photons}
   m_{1+1}=\frac{2\hbar\omega}{c^{\,2}}\sin\vartheta\neq 0.
\end{equation}
This result is easily generalized, for example, for the case of $N$ identical photons moving in one direction and the same number of photon propagating in a different direction, with the angle $2\vartheta$ between these two propagation directions. The mass of such group of $2N$ photons is simply $N$ times larger than that of a single pair of photons (\ref{2photons}):
\begin{equation}
  \label{2Nphotons}
   m_{N+N}=\frac{2N\hbar\omega}{c^{\,2}}\sin\vartheta.
\end{equation}

Note that the defined  in this way Lorentz-invariant mass has nothing in common with a hypothetical nonzero mass of a single photon. Existence of the latter is not proved by any experiments. But, nevertheless, there are works discussing changes in the existing field and  particle theories, which would arise if it would be proved that a single photon has a nonzero mass (see e.g. the papers \cite{Okun3,Lakes, Proca} and references therein).  Nowadays, the existing experiments give only  the upper boundary of the single-photon mass, $m_{single-photon}< 8\times 10^{-51}$ g \cite{Luo}. As this value is extremely small and as, at all, existence of $m_{single-photon}\neq 0$ is very questionable, here and below we remain in the frame of a standard assumption that single photons are massless particles, which does not prevent for groups of photons from having nonzero invariant masses.
Other fields of investigations which are beyond the scope of this work is gravitational interaction of photons and photon beams with other objects \cite{Landau}, and analysis of gravitational field produced by light beams \cite{RWM} . By quoting L.B. Okun \cite{Okun}, ``The mass of a relativistically moving body is not a measure of its inertia. The mass of a relativistically moving body does not determine its interaction with the gravitational field. Despite these "noes" the mass of a body is also an extremely important property in the theory of relativity. A vanishing mass means that the "body" must always move with the speed of light. A nonvanishing mass characterizes the mechanics of a body in a frame of reference in which it is at rest. This frame of reference is distinguished compared with other inertial systems.  According to the theory of relativity, the mass of a particle is a measure of the energy "sleeping" in the particle at rest; it is a measure of the rest energy: $\varepsilon_0 = mc^2$." Of course in these fundamental statements the word "body" has to be understood widely, including for example groups of particles of any kind or groups of bodies.

Classical light fields are believed generally to consist of photons. Usually classical fields are produced in the form of pulses limited in space and time. Such pulses consist of large but finite numbers of photons. Moreover because of diffraction, wave vectors of photons in such pulses are not parallel to each other. For this reason the sets of all photons forming light pulses have nonzero invariant masses, which can be considered as the invariant masses of light pulses. The finiteness of a mass of a diverging light pulse means automatically that the velocity of its propagation in vacuum $v$ is smaller than the light spewed $c$, which agrees with the main conclusion of a recent paper \cite{Padgett} . The main goal of this work is finding explicitly the mass of photons forming light pulses in terms of such parameters of pulses as the electric field strength amplitude, pulse waist and duration, etc. We believe, these tasks are conceptually important. The next two section provide the relevant quantum-electrodynamical (QED) and classical derivations, the mass of classical pulses is found in section 4, the pulse-propagation speed in vacuum is evaluated in section 5, and the existing \cite{Padgett} and other possible experiments are discussed in section 6.

\section{Multimode coherent states}
As known \cite{Glauber}, the best QED  counterpart of a classical field is the so called coherent state
\begin{gather}
 \nonumber
 \ket{\Psi_{{\vec k},\,\sigma}}=e^{-\frac{|\alpha_{{\vec k},\,\sigma}|^2}{2}}\sum_{n_{{\vec k},\,\sigma}}\frac{\alpha_{{\vec k},\,\sigma}^{n_{{\vec k},\,\sigma}}}{\sqrt{n_{{\vec k},\,\sigma}!}}\ket{n_{{\vec k},\,\sigma}}
\\\equiv
  e^{-\frac{|\alpha_{{\vec k},\,\sigma}|^2}{2}}\sum_{n_{{\vec k},\,\sigma}}\frac{(\alpha_{{\vec k},\,\sigma} a_{{\vec k},\,\sigma}^{\dag})^{n_{{\vec k},\,\sigma}}}{n_{{\vec k},\,\sigma}!}\ket{0},
 \label{coh-st}
\end{gather}
where $\ket{0}$ is the vacuum, $\ket{n_{{\vec k},\,\sigma}}$ and  $a_{{\vec k},\,\sigma}^\dag$ are the $n$-photon state vector and the photon creation operator for a given single mode characterized by the photon wave vector ${\vec k}$ and polarization $\sigma$; $\alpha_{{\vec k},\,\sigma}$ are arbitrary complex numbers.

Note that, in accordance with the most often used approach \cite{AB}, the photon modes are defined as plane waves in the periodicity box with the volume $V$, which has to be taken infinitely large in the final results to be derived. In this definition the photon wave vectors ${\vec k}$ are discretized so that, for example, $k_x=2\pi n_x/V^{1/3}$, where $n_x$ is an integer, $n_x=0,1,2,...$, and the same for the $k_y$ and $k_z$. The often met sums over modes can be replaced by integrals over $3D$ wave vectors with the help of the rule
$
\sum_{\vec k}\rightarrow\frac{V}{(2\pi)^3}\int d{\vec k}.
$

By definition, the state vector $\ket{\Psi_{{\vec k},\,\sigma}}$ characterizes the state with an uncertainly large number of photons, such that their wave vectors are identical and parallel to each other. The mass of such formation equals zero. Besides, the electric field strength calculated with the help of the state vector of Eq. (\ref{coh-st}) is a purely plane wave, infinitely extended in space and time and insufficient for describing light pulses. The states appropriate for this goal are the multimode coherent states \cite{Glauber,Wolf}
\begin{equation}
 \ket{\Psi}=\prod_{{\vec k},\,\sigma}
 e^{-\frac{|\alpha_{{\vec k},\,\sigma}|^2}{2}}
 \sum_{n_{{\vec k},\,\sigma}}
  \frac{(\alpha_{{\vec k},\,\sigma}a^\dag_{{\vec k},\,\sigma})^{n_{{\vec k},\,\sigma}}}{n_{{\vec k},\,\sigma}!}\ket{0}.
 \label{MMCoherent}
\end{equation}
Both the single-mode (\ref{coh-st}) and multimode (\ref{MMCoherent}) states obey the normalization condition $\braket{\Psi|\Psi}=1$ and, hence, they can be used for calculating average values of all kinds of operators combined from $a_{{\vec k},\,\sigma}^\dag$ and $a_{{\vec k},\,\sigma}$. In particular, both the single-mode and multimode coherent states can be used for finding the mean number of photons per mode
\begin{equation}
 \label{phot-number}
 \braket{\Psi_{{\vec k},\,\sigma}|a_{{\vec k},\,\sigma}^\dag a_{{\vec k},\,\sigma}|\Psi_{{\vec k},\,\sigma}}=\braket{\Psi|a_{{\vec k},\,\sigma}^\dag a_{{\vec k},\,\sigma}|\Psi}=|\alpha_{{\vec k},\,\sigma}|^2.
\end{equation}
The total number of photons in the multimode coherent state (\ref{MMCoherent}) is given by the sum over modes
\begin{equation}
 \label{phot-number-multi}
 N=\sum_{{\vec k},\,\sigma}\,|\alpha_{{\vec k},\,\sigma}|^2=\frac{V}{(2\pi)^3}
 \int d{\vec k}\,|\alpha_{{\vec k},\,\sigma}|^2.
\end{equation}
Evidently, both the number of photons per mode $|\alpha_{{\vec k},\,\sigma}|^2$ and the total number of photons $N$ are Lorentz-invariant.

The multimode coherent state (\ref{MMCoherent}) can be used also for finding the average energy and momentum of the field
\begin{gather}
 \label{energy-qed}
\braket{\varepsilon} =\sum_{{\vec k}\,\sigma}\hbar\omega_{k}\braket{\Psi|a^\dag_{{\vec k},\sigma}a_{{\vec k},\,\sigma}|\Psi}
=\sum_{{\vec k},\,\sigma}\hbar\omega_{k}|\alpha_{{\vec k},\,\sigma}|^2,\\
 \label{momentum-qed}
 \braket{{\vec p}\,}=\sum_{{\vec k}\,\sigma}{\hbar{\vec k}}\braket{\Psi|a^\dag_{{\vec k},\sigma}a_{{\vec k},\,\sigma}|\Psi}
=\sum_{{\vec k},\,\sigma}{\hbar{\vec k}}\,|\alpha_{{\vec k},\,\sigma}|^2,
\end{gather}
where $\omega_{k}=c|{\vec k}\,|$.

The mass of the state $\ket{\Psi}$ (\ref{MMCoherent}) can be defined via $\braket{\varepsilon}$ and $\braket{{\vec p}\,}$ in the same way (\ref{mass-general}) as for groups of classical particles
\begin{gather}
 \nonumber
 m^2c^4=\braket{\varepsilon}^2 -c^2\braket{{\vec p}\,}^2\\
 \label{massQED}
 =\hbar^2c^2\sum_{{\vec k},{\vec k}^\prime}\Big(k^{(4)}\cdot{k^\prime}^{(4)}\Big)
 \sum_{\sigma,\sigma^\prime}|\alpha_{{\vec k},\,\sigma}|^2|\alpha_{{\vec k}^\prime,\,\sigma^\prime}|^2,
\end{gather}
where $k^{(4)}=\Big(\frac{\omega_{\vec k}}{c},{\vec k}\Big)$ and
${k^\prime}^{(4)}=\Big(\frac{\omega_{{\vec k}^\prime}}{c},{\vec k}^\prime\Big)$ are the photon's 4-momenta (divided by $\hbar$). Once again, both the scalar products of 4-momenta and the numbers of photons in modes $|\alpha_{{\vec k},\,\sigma}|^2$ and $|\alpha_{{\vec k}^\prime,\,\sigma^\prime}|^2$ are Lorentz-invariant, as well as the expressed via them mass of the state $\ket{\Psi}$ (\ref{MMCoherent}). As ${k^{(4)}}^2=0$, evidently, all diagonal terms in the sum over ${\vec k},{\vec k}^\prime$ in Eq. (\ref{massQED}) vanish, whereas the off-diagonal terms determine a nonzero mass of the state $\ket{\Psi}$ (\ref{MMCoherent}).

The electric field strength is defined as the averaged value of its operator expression, $\braket{\Psi|{\hat E_\sigma}|\Psi}$, where
\begin{gather}
 \nonumber
 {\hat E}_\sigma=i\sum_{\vec k} \sqrt{\frac{2\pi\hbar\omega_k}{V}}\times\\
 \label{E-operator}
 \left[a_{{\vec k},\,\sigma}
 e^{i({\vec k}{\vec r}-\omega_kt)}-a^\dag_{{\vec k},\,\sigma}
 e^{-i({\vec k}{\vec r}-\omega_kt)}\right],
\end{gather}
which gives
\begin{gather}
 \nonumber
 \braket{E}_\sigma({\vec r},t)=i\sum_{\vec k} \sqrt{\frac{2\pi\hbar\omega_k}{V}}\,\,|\alpha_{{\vec k},\,\sigma}|\times\\
 \label{E-av-QED}
 \left[e^{i({\vec k}{\vec r}-\omega_kt+\varphi_{{\vec k},\,\sigma})}-e^{-i({\vec k}{\vec r}-\omega_kt+\varphi_{{\vec k},\,\sigma})}\right],
\end{gather}
where $\varphi_{{\vec k},\,\sigma}$ is the phase of $\alpha_{{\vec k},\,\sigma}$.

With the sums over ${\vec k}$ in Eqs. (\ref{energy-qed}), (\ref{momentum-qed}) and (\ref{E-av-QED}) transformed into integrals, these equations are reduced to the form
\begin{gather}
 \label{energy-qed-int}
\braket{\varepsilon} =\sum_\sigma\frac{V}{(2\pi)^3}\int d{\vec k}\,
\hbar\omega_{k}\,|\alpha_{{\vec k},\,\sigma}|^2\\
 \label{momentum-qed-in}
 \braket{{\vec p}\,}=\sum_\sigma\frac{V}{(2\pi)^3}\int d{\vec k}\,
\hbar{\vec k}\,|\alpha_{{\vec k},\,\sigma}|^2,
\end{gather}
and
\begin{gather}
 \nonumber
 \braket{E}_\sigma({\vec r},t)=\frac{1}{2^{3/2}\pi^{5/2}}\int d{\vec k}\, \sqrt{V\hbar\omega_k}\,\,|\alpha_{{\vec k},\,\sigma}|\times\\
 \label{E-av-QED-int}
 \sin(\omega_kt-{\vec k}{\vec r}-\varphi_{{\vec k},\,\sigma}).
\end{gather}

The next step consists in comparison with the Fourier representation of the field describing light pulses.

\section{Classical description}

Let us consider the boundary problem with the field propagating along the $z$-axis and defined in the half-space $z>0$ by its distribution in the $(xy)$ plane at $z=0$:
\begin{equation}
 \label{at z=0}
 E_\sigma({\vec r}_\perp,t)|_{z=0}=E_{\sigma}^{(0)}({\vec r}_\perp)\sin(\omega_0t)\,e^{-t^2/2\tau^2},
\end{equation}
where $\tau$ and $\omega_0$ are the pulse duration and carrier frequency of the field, and the temporal shape of the pulse is taken Gaussian. As for the dependence of the field amplitude on transverse coordinates, at this stage it can be left unspecified, though in final results it will be taken Gaussian too. But in any case, $E_\sigma({\vec r}_\perp,t)|_{z=0}$ is assumed to be real. The expression on the right-hand side of Eq. (\ref{at z=0}) can be Fourier transformed both with respect to $t$ and ${\vec r}_\perp$ to be reduced to the form
\begin{gather}
 \nonumber
 E_\sigma({\vec r}_\perp,t)|_{z=0}=\frac{i\tau\quad}{2(2\pi)^{3/2}}\int d{\vec k}_\perp  d\omega\, e^{i{\vec k}_\perp \cdot{\vec r}_\perp-i\omega t}{\widetilde E}_{\sigma}^{(0)}({\vec k}_\perp)\\
 \label{at z=0-Fourier}
 \times\left[e^{-(\omega+\omega_0)^2\tau^2/2}-e^{-(\omega-\omega_0)^2\tau^2/2}\right],
\end{gather}
where ${\widetilde E}_{\sigma}^{(0)}({\vec k}_\perp)$ is the Fourier transform of the transverse field envelope at $z=0$, $E_{\sigma0}^{(0)}({\vec r}_\perp)$
\begin{equation}
 \label{Fourier}
 {\widetilde E}_{\sigma}^{(0)}({\vec k}_\perp)=\frac{1}{2\pi}\int d{\vec r}_\perp e^{-i{\vec k}_\perp \cdot{\vec r}_\perp} E_{\sigma}^{(0)}({\vec r}_\perp)
\end{equation}
such that ${\widetilde E}_{\sigma}^{(0)}(-{\vec k}_\perp)=\left[{\widetilde E}_{\sigma}^{(0)}({\vec k}_\perp)\right]^*$.

By changing sign of the integration variable $\omega$ in the integral containing the first term in square brackets in the last line of Eq. (\ref{at z=0-Fourier}), we can reduce the whole integral over $\omega$ to a more convenient form
\begin{equation}
 \label{in-over-omega}
 \int d\omega\,e^{-(\omega-\omega_0)^2\tau^2/2}\left[e^{i\omega t}-e^{-i\omega t}\right].
\end{equation}
Now a simple substitution $\omega t\rightarrow \omega t-k_z z$ with $k_z=\sqrt{\frac{\omega^2}{c^2}-{{\vec k}_\perp}^2}$ gives the field distribution in the whole half-space $z>0$, with the field obeying Maxwell equations
\begin{gather}
 \nonumber
 E_\sigma({\vec r}_\perp,t,z)=i\frac{\tau}{2(2\pi)^{3/2}}\int d{\vec k}_\perp \int d\omega\, e^{i{\vec k}_\perp \cdot{\vec r}_\perp}{\widetilde E}_{\sigma}^{(0)}({\vec k}_\perp)\\
 \label{at z neq 0}
 \times e^{-\frac{(\omega-\omega_0)^2\tau^2}{2}}\left[e^{-i(\omega t-k_z z) }-e^{+i(\omega t-k_z z)}\right].
\end{gather}
At last, the integration over $\omega$ can be substituted by integration over $k_z$ with the integral transformation rule $\int d\omega=\int dk_z \frac{c^2k_z}{\omega_k}$, where $\omega_k=ck=c\sqrt{k_z^2+{\vec k}_\perp^2}$. Transformed in this way, the expression for the field strength (\ref{at z neq 0}) takes the form
\begin{gather}
 \nonumber
 E_\sigma({\vec r},t)=\frac{\tau}{(2\pi)^{3/2}}\int d{\vec k}\, \left|{\widetilde E}_{\sigma}^{(0)}({\vec k}_\perp)\right|\,\frac{c^2k_z}{\omega_k} e^{ -(\omega_k-\omega_0)^2\tau^2/2}\\
 \label{final}
 \times\,\sin\left[\omega_k t-{\vec k}{\vec r}-\varphi_\sigma({\vec k}_\perp)\right],
\end{gather}
where $\varphi_\sigma({\vec k}_\perp)$ is the phase of the function ${\widetilde E}_{\sigma}^{(0)}({\vec k}_\perp)$ (\ref{Fourier}).

Note that Eq. (\ref{final}) represents a special case of the well known \cite{Landau} general rule for the expansion of an arbitrary-configuration light field $E({\vec r},t)$ in  a series (integral) of field eigenfunctions obeying Maxwell equations
\begin{equation}
 \label{expans-general}
 E({\vec r},t)=\frac{1}{(2\pi)^{3/2}}\int d{\vec k}\left[{\widetilde E}_{\vec k}\,e^{i({\vec k}{\vec r}-\omega t)}+{\widetilde E}^*_{\vec k}\,e^{-i({\vec k}{\vec r}-\omega t)}\right],
\end{equation}
with
\begin{equation}
 \label{expans-general}
{\widetilde E}_{\vec k}\,e^{-i\omega_k t}+{\widetilde E}^*_{-\vec k}\,e^{i\omega_k t} =\frac{1}{(2\pi)^{3/2}}\int d{\vec r}\,E({\vec r},t)e^{-i{\vec k}{\vec r}}.
\end{equation}

\section{The mass of a light pulse}

By comparing two expressions for the light field strength, (\ref{E-av-QED-int}) and (\ref{final}), we find that the coherent-state parameters $\alpha_{{\vec k},\,\sigma}$ are related directly to the Fourier transformed classical field strength of a light pulse
\begin{equation}
 \label{alpha-E}
 |\alpha_{{\vec k},\,\sigma}|=\frac{\pi\tau}{\sqrt{V\hbar\omega_k}}\,|{\widetilde E}_{\sigma}^{(0)}({\vec k}_\perp)|\frac{c^2|k_z|}{\omega_k} e^{ -(\omega_k-\omega_0)^2\tau^2/2},
\end{equation}
with $\varphi_{{\vec k},\,\sigma}=\varphi_\sigma({\vec k}_\perp)$. By substituting $|\alpha_{{\vec k},\,\sigma}|$ of Eq. (\ref{alpha-E}) into the QED equations (\ref{energy-qed-int}) and (\ref{momentum-qed-in}) we get absolutely classical expressions for the mean (total) energy and momentum
\begin{gather}
 \label{energy-final}
 \braket{\varepsilon}=\frac{(c\tau)^2}{8\pi}\int d{\vec k}\,
\,|{\widetilde E}^{(0)}({\vec k}_\perp)|^2\Big(\frac{ck_z}{\omega_k}\Big)^2 e^{ -(\omega_k-\omega_0)^2\tau^2}
\end{gather}
and
\begin{gather}
 \label{momentum-final}
 \braket{{\vec p}\,}=\frac{(c\tau)^2}{8\pi}\int d{\vec k}
\,|{\widetilde E}^{(0)}({\vec k}_\perp)|^2\frac{{\vec k}}{\omega_k}\Big(\frac{ck_z}{\omega_k}\Big)^2 e^{ -(\omega_k-\omega_0)^2\tau^2}.
\end{gather}
Note that the dependence of the electric field strength on the polarization index $\sigma$ is determined by projections $e_\sigma$ on the $\sigma$-axes of the field polarization unit vector ${\vec e}_{\vec k}$ (such that ${\vec e}_{\vec k}\perp{\vec k}$). When the squared absolute values of the field amplitudes ${\widetilde E}_\sigma^{(0)}$ are summed over $\sigma$, their polarization parts turn unit: $\sum_\sigma |e_\sigma|^2\equiv |{\vec e}_{\vec k}|^2=1$. This explains why summations over and any dependencies on $\sigma$ disappear in  Eqs. (\ref{energy-final}), (\ref{momentum-final}) and in all formulas below.

The mass of the pulse is defined by a standard relation of the type (\ref{mass-general}), which gives
\begin{equation}
 \label{mass-pulse}
 m^2c^4=\Big(\braket{\varepsilon}+|c\braket{{\vec p}\,}|\Big)\Big(\braket{\varepsilon}-|c\braket{{\vec p}\,}|\Big).
\end{equation}

Let us assume now that both the transverse distribution of the field in the plane $z=0$ and its Fourier transform are Gaussian
\begin{equation}
 \nonumber
 E^{(0)}({\vec r}_\perp)=E_0 e^{-{\vec r}_\perp^2/2w^2},\\
 \label{Gaussian transv}
 \widetilde{E}^{(0)}({\vec k}_\perp)=E_0w^2 e^{-{\vec k}_\perp^2 w^2/2},
\end{equation}
where $w$ is the pulse width (waist) at $z=0$. Because of the axial symmetry of these distributions, Eq. (\ref{momentum-final}) gives immediately $\braket{p_x}=\braket{p_y}=0$, and there is only one non-zero $z$-component of the total pulse momentum, $\braket{p_z}\neq 0$. Let us assume also that, as usual, the length and transverse size of the pulse exceed significantly its wavelength
\begin{equation}
 \label{conditions}
 w,\,c\tau\gg\lambda=\frac{2\pi c}{\omega_0}.
\end{equation}
Under these conditions the exponential factor in Eqs. (\ref{energy-final}) and (\ref{momentum-final}) takes the form
\begin{equation}
 \label{exp}
 e^{-(\omega_k-\omega_0)^2\tau^2}\approx e^{-(k_z-\omega_0/c)^2(c\tau)^2},
\end{equation}
and the factor $(ck_z/\omega_k)^2$ can be approximated by unit to give
\begin{gather}
 \nonumber
 \braket{\varepsilon}\approx c\braket{p_z}=\frac{E_0^2(c\tau)^2}{8\pi}\int d{\vec k}\,
 e^{-{\vec k}_\perp^2w^2}e^{ -(k_z-\omega_0/c)^2(c\tau)^2}\\
 \label{en=mom}
 =\frac{\sqrt{\pi}c\tau w^2 E_0^2}{8}
\end{gather}
and
\begin{equation}
 \braket{\varepsilon}+c\braket{p_z}\approx
 \label{sum}
 \frac{\sqrt{\pi}c\tau w^2 E_0^2}{4}.
\end{equation}
As for the difference between $\braket{\varepsilon}$ and $c\braket{p_z}$ in the definition of mass (\ref{mass-pulse}), there is a rather strong compensation of these two terms because of the arising in this difference factor
$$1-\frac{ck_z}{\omega_k}\approx \frac{{\vec k}_\perp^2}{2(\omega_0/c)^2}\,,$$ which gives
\begin{gather}
 \nonumber
 \braket{\varepsilon}-c\braket{p_z}=\frac{E_0^2(c\tau)^2}{16\pi(\omega_0/c)^2}\int d{\vec k}\,{{\vec k}_\perp}^2
 e^{-{{\vec k}_\perp}^2w^2}e^{ -(k_z-\omega_0/c)^2(c\tau)^2}\\
 \label{diff}
 =\frac{\sqrt{\pi}c\tau E_0^2}{16(\omega_0/c)^2}.
\end{gather}
Substituted into the definition of mass (\ref{mass-pulse}), Eqs. (\ref{sum}) and (\ref{diff}) give the following final result:
\begin{equation}
 \label{massa}
 m=\frac{\sqrt{\pi}\tau w E_0^2}{8\omega_0}=\frac{1}{16\sqrt{\pi}}\frac{E_0^2\tau w\lambda}{c}.
\end{equation}
Comparison with the energy $\braket{\varepsilon}$ (\ref{en=mom}) gives
\begin{equation}
 \label{comparison}
 m=\frac{\braket{\varepsilon}}{2\pi c^2}\frac{\lambda}{w}\ll \frac{\braket{\varepsilon}}{c^2}.
\end{equation}

Numerically, for example, $m=10^{-20}$ g in the case of a pulse with energy 10 mJ (intensity $10^{10}$ W/cm$^2$ and pulse duration 1 ps), pulse waist $w=1$ cm, and wave length $\lambda=1 \mu$m.

For the same parameters, the number of photons in a pulse is
\begin{equation}
 \label{phot-numb}
 N=\frac{\braket{\varepsilon}}{\hbar\omega_0}\approx 10^{17}.
\end{equation}
In terms of the photon number $N$ the expression (\ref{massa}) for the pulse invariant mass takes the form
\begin{equation}
 \label{mass-via-N}
 m=\frac{N\hbar\omega_0}{2\pi c^2}\frac{\lambda}{w}.
\end{equation}
This result is similar to that occurring for two beams of photons with $N$ collinear photons in each beam and angle $\vartheta$ between the propagation directions of beams. (\ref{2Nphotons}). This similarity turns into total coincidence if the angle $\vartheta$ is taken equal to the diffraction divergence angle in a beam with appropriately chosen numerical factor, $\vartheta = \lambda/4\pi w\ll 1$. This means that the model of two beams of photons describes qualitatively, but sufficiently well, features of the diverging classical beam of light.

Another interesting point is a transition to an infinitely wide beam, $w\rightarrow\infty$. The first impression is that in this limit the beam turns into a plane wave in which all photons are collinear and the invariant mass has to turn zero.  But, in fact, this is not as evident as it seems to be.  The result depends on what is changing and what remains constant when $w\rightarrow\infty$. If the transverse size of a light beam $w$ grows at constant values of the field strength amplitude $E_0$ and pulse duration $\tau$, then, in accordance with Eq. (\ref{massa}), the mass $m$ grows, $m|_{E_0={\rm const.}}\propto w\rightarrow\infty$ at $w\rightarrow\infty$. Oppositely, if $w$ grows at a given constant number of photons $N$ in a pulse-volume (or energy in a pulse $\braket{\varepsilon}$), then, as follows from  Eqs. (\ref{comparison}) and  (\ref{mass-via-N}), the mass of the pulse falls, $m|_{N={\rm const.}}\propto 1/w\rightarrow 0$ at $w\rightarrow\infty$. And, of course, there are many intermediate cases, when both $E_0$ and $N$ are varying in some ways with varying transverse size of  a light beam $w$, and the mass  $m|_{w\rightarrow\infty}$ can take any value between $0$ and $\infty$  depending on the form of the functions $E_0(w)$  and $N(w)$.

As a final remark to this section, it's worth mentioning that in our recent paper \cite{LPL} we have introduced the so called Lorentz-invariant density of mass of a light field defined as the squared density of energy minus the squared Poynting vector
\begin{equation}
 \label{mu}
 \mu({\vec r}, t)=\left(\frac{E^2+H^2}{8\pi c}\right)^2-\left(\frac{{\vec E}\times{\vec H}}{4\pi c}\right)^2=inv.,
\end{equation}
 where ${\vec E}({\vec r}, t)$ and ${\vec H}({\vec r}, t)$ are the electric and magnetic field strengths. Invariance of $\mu({\vec r}, t)$ is proved by its identical expression in terms of the field invariants, $\mu=\frac{1}{8\pi c^2}\left[(E^2-H^2)^2+4({\vec E}\cdot{\vec H})^2\right]^{1/2}$. Under the name of {\it electromagnetic inertia density} the same value as $\mu({\vec r}, t)$ was introduced earlier and is discussed nowadays by G. Kaiser \cite{Kaiser} (see also \footnote[18]{ G. Kaiser,  Completing the complex Poynting theorem: Conservation of reactive energy in reactive time, arXiv:1412.3850v2 [math-ph], 2016}). Concerning the results of our present investigation, evidently, the invariant mass of a light pulse $m$ (\ref{massa}), (\ref{comparison}), (\ref{mass-via-N}) cannot be interpreted as the invariant mass density $\mu({\vec r}, t)$ integrated over the pulse volume because, clearly, $\int\mu d{\vec r}$ is not Lorentz-invariant. Thus, these two concepts are different and are not connected directly with each other. They characterize different features of light pulses: if the mass density $\mu({\vec r}, t)$ characterizes local features of the field in a pulse, the mass $m$ characterizes global features of the pulse as a whole.

\section{Propagation speed and the "rest frame" for light pulses}

\begin{figure}[h]
\centering\includegraphics[width=8 cm]{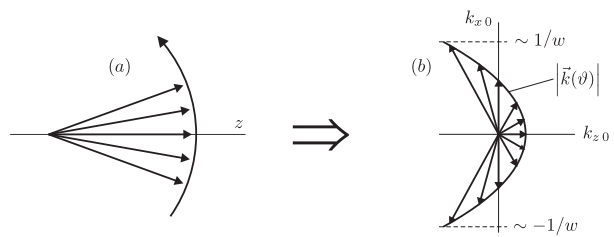}
\caption{{\protect\footnotesize {Distribution of photon wave vectors of a pulse $(a)$ in laboratory and $(b)$ rest frames.}}}\label{Fig1}
\end{figure}

 In accordance with the general relativistic formulas (\ref{mass1}) and (\ref{velocity}), the propagation speed of a pulse is determined as $v=c^2\braket{p_z}/\braket{\varepsilon}$ which gives
\begin{equation}
 v=c\left(1-\frac{\braket{\varepsilon}^2-c^2\braket{p_z}^2}
 {\braket{\varepsilon}(\braket{\varepsilon}+c\braket{p_z})}\right)
 \label{v}
 \approx c\left(1-\frac{m^2c^4}{2\braket{\varepsilon}^2}\right)
\end{equation}
or
\begin{equation}
 \label{c-v}
 c-v\approx c\,\frac{m^2c^4}{2\braket{\varepsilon}^2}=\frac{c}{8\pi^2}\frac{\lambda^2}{w^2}.
\end{equation}
This result shows that the propagation speed of light pulses $v$ in vacuum is smaller than the light speed $c$, which agrees with the conclusion of the recent paper \cite{Padgett}. Moreover, Eq. (\ref{v}) shows that the difference between $c$ and $v$ is determined just by the squared Lorentz-invariant mass of pulses. This observation shows that, in principle, the mass of pulses can be measured directly by means of measurement of the pulse-propagation speed. Next, as the pulse-propagation speed $v$ (\ref{v}) is less than the speed of light, in the  inertial frame moving together with the pulse, i.e. with the speed $v$ with respect to the laboratory frame, the pulse will look as stopped, with completely eliminated  translatory motion.
Indeed, as easily checked, in this frame the pulse momentum equals zero
\begin{equation}
 \label{zer0-pz-rf}
 \braket{p_z}_0=\left.\frac{\braket{p_z}-v\braket{\varepsilon}/c^2}
 {\sqrt{1-v^2/c^2}}\right|_{v=c^2\braket{p_z}/\braket{\varepsilon}}=0,
\end{equation}
where the subscript ``0" indicates the rest-frame for a light pulse. In this frame the center of mass of a pulse does not move, and the only motion remaining in a pulse is its spreading. Spreading can be considered as an internal motion in a system consisting of photons forming a light pulse. The distribution of individual wave vectors of photons in the pulse in the rest frame is shown on the right-hand side of Fig. \ref{Fig1}, and it looks rather peculiar.
If in the laboratory frame orientation of wave vectors is determined by their angle $\vartheta$ with respect to the $z$-axis, in the rest frame components of wave vectors are given by the usual Lorentz-transformation formulas
\begin{equation}
 \label{Lor-transf}
 k_{z\, 0}(\vartheta)=\gamma\omega\ (\cos\vartheta-v/c),  k_{z\, 0}(\vartheta)=\omega\sin\vartheta,
\end{equation}
where $\gamma=\left[1-v^2/c^2\right]^{-1/2}$. The solid line in the picture on the right-hand side of Fig. \ref{Fig1} indicates ending of wave vectors starting from the origin and determined by the angle $\vartheta$ as a parameter. As follows from the first formula of Eq. (\ref{Lor-transf}), at $\vartheta=0$ the $z$-component of the wave vector $k_{z\, 0}$ is always positive. This means that there is no inertial frame moving along the $z$-axis in which the direction of the wave vector $k_{z\, 0}(\vartheta)$ could be inverted. This is possible only for wave vectors with relatively large values of $\vartheta$. As whole, in the rest frame, the sum of positive and negative $z$-components of wave vectors compensate each other to give $\sum k_{z\, 0}=0$.

In the rest-frame the pulse energy $\braket{\varepsilon}_0$ equals its invariant mass multiplied by the squared speed of light,
\begin{equation}
 \label{rest-frame}
 \braket{\varepsilon}_0=mc^2.
\end{equation}
This is the main physical meaning of the invariant mass as a quantity determining internal energy of a system under consideration not related to its translatory motion. In application to light pulses, the definitions of invariant mass and the rest-frame energy are equivalent, and one of them can be used instead of another. On the other hand, a real transition to the frame moving with the pulse speed can be problematic, whereas measurement of the pulse-propagation speed itself, in principle, is doable. With results of such measurement known, one can use Eq. (\ref{v}) for finding the invariant mass (\ref{massa}) and, then, Eq. (\ref{rest-frame}) for finding a proper energy of the pulse in the rest frame. As in the laboratory frame the mass is $\lambda/w$ times smaller than the pulse energy divided by $c^2$ (\ref{comparison}) , the same is true for the rest-frame energy
 \begin{equation}
  \label{comp-rf}
 \braket{\varepsilon}_0=\frac{\lambda}{2\pi w}\braket{\varepsilon}_{lab}\ll\braket{\varepsilon}_{lab}.
\end{equation}
The pulse energy in the rest-frame is minimal compared to the pulse energy in any other frames, and the pulse energy in the laboratory frame is determined mainly by the pulse propagation or translatory motion rather than by its spreading.

\section{Experiments}
In a very interesting experimental work \cite{Padgett}, slowing down of photon propagation in vacuum was observed  in a scheme of Spontaneous Parametric Down Conversion (SPDC). Pairs of photons were split for two channels. In one of them photons were propagating in a fiber with a given velocity close to the speed of light in vacuum. In the second channels photons propagated in a free space. Owing to the presence of a transverse component in their wave vectors, the velocity of propagation of these photons along the pump-propagation axis was smaller than the speed of light. This effect was registered by the measured positions of the minimum in the Hong-Ou-Mandel's curve of the coincidence signal vs the delay time of the free-space propagating photons \cite{HOM}. This position of the minimum was found to be dependent on  the transverse component ${\vec k}_\perp$ of the total momentum of free-space propagating photons. The velocity of photon propagation along the pump-propagation axis was found \cite{Padgett} to be given by
\begin{equation}
 \label{SPDC-speed}
  v=\left(1-\frac{k_\perp^2}{2\left|{\vec k}\right|^2}\right).
 \end{equation}
Authors of the work \cite{Padgett} do not interpret their results in terms of the invariant mass of SPDC photons though, undoubtedly, such interpretation is possible. Moreover, by comparing expressions of Eqs. (\ref{v}) and (\ref{SPDC-speed}) we find that there is a very simple one-to-one correspondence between the invariant mass and momentum (wave vector) of transverse motion:
\begin{equation}
 \label{m-k-perp}
  \frac{\braket{k_\perp^2}}{2\left|{\vec k}\right|^2}\equiv\frac{m^2c^4}{2\varepsilon^2}.
\end{equation}

\begin{figure}[h]
\centering\includegraphics[width=8 cm]{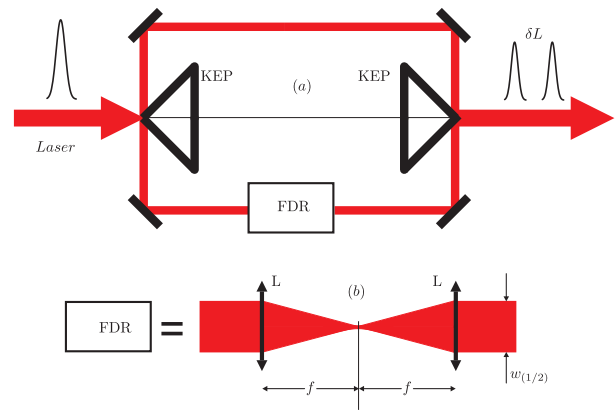}
\caption{{\protect\footnotesize {A suggested scheme for observation of the light-pulse slowing down owing to its focusing-defocusing. $(a)$ KEP are the knife edge prisms, $\delta L$ is the distance between the slowed down and non-slowed pulses, FDR is the focusing-defocusing region, $\delta L$ is the spacing betwee4n the slowd down and non slowed pulses; $(b)$ light pulse in the focusing-defocusing region, L denotes confocal lenses, $f$ is the focus distance of each of them, $w_{(1/2)}$ is the waist of the pulse outside of the  focusing- defocusing region.}}}\label{Fig2}
\end{figure}

As for direct measurements of the slowing down effect in the case of real Gaussian laser pulses, the main problem is in a very small value of the invariant mass (\ref{massa}), (\ref{comparison}), (\ref{mass-via-N}) and relatively very small difference of velocities $c-v$ (\ref{c-v}). But the effect can be significantly  strengthened by means of focusing and defocusing of the light pulse. The scheme of an experiment we suggest is shown in Fig. \ref{Fig2}. In this scheme some ideas of the experiment \cite{Padgett} are used and adapted to the case of laser pulses.
We assume that the laser beam has originally a rather large waist, e.g., $w\sim 1\, cm$, which corresponds to a very weak  diffraction divergence and a very large diffraction length, $L_D\sim w^2/\lambda\sim 10^4\, cm$ at $\lambda= 1\mu m$. Similarly to the experiment \cite{Padgett}, in the scheme of Fig. \ref{Fig2} the laser beam is supposed to be split by a knife edge prism for two parts with two half-pulses propagating one in the upper and one in the lower channels. Let their transverse sizes are $w_{(1/2)}\sim w/2$. The pulse in the upper channel experiences no additional changes whereas the the pulse in the lower channel is assumed to be focused by a lens with the focal length $f$ and then, at the distance $2f$ from the first lens, it is returned to its original form with the waist $w_{(1/2)}$ and with a very weak diffraction divergence. After the focusing and defocusing in the lower channel both pulses are sent again the knife edge prism to merge into a single beam. If the focusing-defocusing region provides any slowing down effect, pulses coming from the lower channel will be delayed with respect to pulses coming from the upper channel. If the delay length is sufficiently pronounced to exceed duration of the original laser pulse, at the exit form the set up one will see two well separated pulses instead of a single one. This will be a clear and direct  evidence that the propagation velocity in the lower channel is less than the the speed of light $c$.

Convergence of the pulse before the focus and its divergence after the focus correspond to appearance of relatively large transverse components of photon wave vectors
\begin{equation}
 \label{k-perp}
 |k_\perp|\sim k_{\perp\,{\rm foc}}\approx \frac{w_{(1/2)}}{f}\,|{\vec k}|,
\end{equation}
as long as $w_{(1/2)}\ll f$. The pulse-propagation velocity in the region between two lenses in Fig. \ref{Fig2} can be defined by Eq. (\ref{SPDC-speed}) with $k_\perp^2$ substituted by $\braket{k_\perp^2}$. If the distribution function $F(k_\perp)$ of a focused beam is taken Gaussian, $F(k_\perp)\propto\exp(-k_\perp^2/k^2_{\perp\,{\rm foc}})$, the average squared transverse wave vector $\braket{k_\perp^2}$ coincides with $k^2_{\perp\,{\rm foc}}$. As a result, we find the following expression for the pulse-propagation velocity in the region between two lenses
\begin{equation}
 \label{vel-in-fig2}
 v=c\left[1-\frac{1}{2}\left(\frac{w_{(1/2)}}{f}\right)^2\right].
\end{equation}
The total spacial delay of the pulse accumulated in the whole region between two lenses is given by
\begin{equation}
 \label{delta l}
 \delta L=2f\left(1-\frac{v}{c}\right)=\frac{w_{(1/2)}^2}{f}.
\end{equation}
For example, at $w_{(1/2)}=0.5\, {\rm cm}$ and $f=5 \, {\rm cm}$, Eq. (\ref{delta l}) gives $\delta L=0.05\, {\rm cm}=0.5\,{\rm mm}$. This is a rather high value exceeding significantly length of laser pulses both of femto and picosecond durations. This shows that the effect of slowing down the propagation speed of laser pulses in vacuum (air) is observable.

The difference of velocities $c-v$ determined by Eq. (\ref{vel-in-fig2}) exceeds significantly the same difference related to the proper diffraction divergence of the beam (\ref{c-v}) if $w_{(1/2)}/f\gg \lambda/w_{(1/2)}$ if
$$
\frac{w_{(1/2)}}{f}\gg\frac{\lambda}{2\pi w_{(1/2)}}\quad{\rm or}\quad f\ll 2\pi\frac{w_{(1/2)}^2}{\lambda}\equiv L_D,
$$
i.e., if the focal length of lenses $f$ is much shorter than the diffraction length of the original light beam $L_D$.

As the propagation velocity and invariant mass of pulses are related to each other by Eq.(\ref{c-v}), the described decrease of the propagation velocity in the focusing-defocusing region is accompanied by increase of the invariant mass $m$ which becomes equal to
\begin{equation}
 \label{mass-fdr}
 m_{\rm fdr}=\frac{\varepsilon}{c^2}\frac{w_{(1/2)}}{f}.
\end{equation}

Note that in principle, there is another very simple and trivial way of getting a double-peak structure at exit of set up like that shown in Fig. 2$a$ by means of lengthening the pathlength  in the lower channel. The method considered here is qualitatively different. The pathlength in the lower channel is not modified in any way and remains equal to the pathlength in the upper channel. What is done by the focusing-defocusing region is the temporal spacial restructuring of the lower-channel light pulse, and this restructuring is the factor that slows down the pulse propagation.

\section{Conclusion}

To resume, the concept of nonzero invariant mass is shown to be appropriate for characterizing features of diffracting light pulses. Its explicit expression is found in terms of laser-pulse parameters: pulse energy, duration, waist and wavelength. The invariant mass of pulses is shown to be related directly with the propagation velocity of pulses in vacuum, which is found to be smaller than the light speed in all cases except an infinitely extended plane wave. It's shown also that the effects of slowing down light pulses and increasing their invariant mass can be strengthened significantly by focusing or defocusing. Based on this, a scheme with two confocal lenses is proposed for measuring directly the pulse-propagation velocity and its deviation from the speed of light. The experiment is found to be doable with femto- and even picosecond pulses. If measured, the pulse-propagation velocity can be used for finding experimentally the invariant mass of a pulse under consideration.

\section*{Acknowledgement}
The work was supported by the grant RFBR 14-02-00811.


\bibliography{text}

\begin{thebibliography}{10}

\bibitem{Landau}
L.~D. Landau and E.~M. Lifshits.
\newblock {\em The Classical Theory of Fields, Fourth Edition}.
\newblock Butterworth-Heinemann, Oxford, UK, 1980.

\bibitem{Okun}
L.~B. Okun.
\newblock The concept of mass (mass,energy, relativity).
\newblock {\em Sov. Phys.Usp.}, 32(7):629 -- 638, 1989.

\bibitem{Okun1}
L.~B. Okun.
\newblock Reply to the letter 'what is mass?' by R I Khrapko.
\newblock {\em Physics Uspekhi}, 43(12):1270 -- 1275, 2000.

\bibitem{Okun2}
L.~B. Okun.
\newblock {\em Energy and Mass in Relativity Theory}.
\newblock World Scientific, Singapore, 2008.

\bibitem{Rivlin}
L.~V. Rivlin.
\newblock Is the photon mass zero?
\newblock {\em Sov. J. Quantum Electron.}, 22(8):771 -- 773, 1992.

\bibitem{Okun3}
I.~Yu. Kobzarev and L.~B. Okun.
\newblock On the photon mass.
\newblock {\em Sov. Phys. Usp.}, 11(3):338 --341, 1968.

\bibitem{Lakes}
R.~Lakes.
\newblock Experimental limits on the photon mass and cosmic magnetic vector potential.
\newblock {\em Phys. Rev. Lett.}, 80(3):1826--1829, 1998.

\bibitem{Proca}
A.~Proca.
\newblock Free particles, photons and particles of 'pure charge'.
\newblock {\em Journal de Physique et le Radium}, 8(1):23--28, 1937.

\bibitem{Luo}
Jun~Luo et~al.
\newblock Experimental limits on the photon mass and cosmic magnetic vector
  potential.
\newblock {\em Phys. Rev. Lett.}, 90(3):081801, 2003.

\bibitem{RWM}
D.~Ratzel, M.~Wilkens, and R.~Menzel.
\newblock Gravitational properties of light - the gravitational field of a laser pulse.
\newblock {\em New J. Phys}, 18(2):023009, 2016.

\bibitem{Padgett}
D.~Giovannini et~al.
\newblock Spatially structured photons that travel in free space slower than the speed of light.
\newblock {\em Science}, 347(6224):857--860, 2015.

\bibitem{Glauber}
R.~J. Glauber.
\newblock Coherent and incoherent states of the radiation field.
\newblock {\em Phys. Rev.}, 131(6):2766 -- 2778, 1963.

\bibitem{AB}
A.~I. Akhiezer and V.~B. Berestetski.
\newblock {\em Quantum Electrodynamics}.
\newblock John Wiley and Sons Inc, Oxford, UK, 1965.

\bibitem{Wolf}
L.~Mandel and E.~Wolf.
\newblock {\em Optical Coherence and Quantum Optics}.
\newblock Cambridge University Press, New Jersey, USA, 1995.

\bibitem{LPL}
S.~V. Vintskevich, V.~G Veselago, and M~V Fedorov.
\newblock On a possible definition of the concept of mass density for a
  classical electromagnetic field in vacuum.
\newblock {\em Laser. Phys. Lett.}, 12(12):096201, 2015.

\bibitem{Kaiser}
G.~Kaiser.
\newblock Electromagnetic inertia, reactive energy and energy flow velocity.
\newblock {\em J. Phys. A: Math. Theor.}, 44(7):345206, 2011.

\bibitem{HOM}
C.K. Hohg, Z.Y. Ou, and L.~Mandel.
\newblock Measurement of subpicosecond time intervals between two photons by interference.
\newblock {\em Phys. Rev. Lett.}, 59(18):2044--2046, 1987.

\end{thebibliography}
%
%

\end{document}